\documentclass[sigconf,pbalance]{acmart}
\AtBeginDocument{%
  }

\copyrightyear{2025}
\acmYear{2025}
\setcopyright{rightsretained}
\acmConference[SCA '25]{SupercomputingAsia 2025}{March 10--13, 2025}{Singapore, Singapore}
\acmBooktitle{SupercomputingAsia 2025 (SCA '25), March 10--13, 2025, Singapore, Singapore}
\acmPrice{}
\acmDOI{10.1145/3718350.3718359}
\acmISBN{979-8-4007-1250-0/25/03}

\usepackage{listings}
\usepackage{calc}
\usepackage{enumitem}
\usepackage[ruled, lined, longend, linesnumbered]{algorithm2e}
\usepackage{algorithmic}
\usepackage{amsmath}
\usepackage{subcaption}
\usepackage{stmaryrd}
\usepackage[nolist,nohyperlinks]{acronym}
\usepackage{soul}
\usepackage{balance}
\usepackage{graphicx}

\definecolor{codegreen}{rgb}{0,0.6,0}
\definecolor{codegray}{rgb}{0.5,0.5,0.5}
\definecolor{codepurple}{rgb}{0.58,0,0.82}
\definecolor{backcolour}{rgb}{0.95,0.95,0.92}

\begin{acronym}
    \newacro{a2c}[A2C]{advantage actor-critic}
    \newacro{cdf}[CDF]{cumulative distribution function}
    \newacro{cl}[CL]{curriculum learning}
    \newacro{drl}[DRL]{deep \ac{rl}}
    \newacro{dnn}[DNN]{deep neural networks}
    \newacro{fcfs}[FCFS]{first-come-first-serve}
    \newacro{hpc}[HPC]{high-performance computing}
    \newacro{mdp}[MDP]{Markov decision process}
    \newacro{ml}[ML]{machine learning}
    \newacro{pdf}[PDF]{probability density function}
    \newacro{psmp}[PSMP]{power state management problem}
    \newacro{qos}[QoS]{quality of service}
    \newacro{rmse}[RMSE]{root mean squared error}
    \newacro{rl}[RL]{reinforcement learning}
    \newacro{swf}[SWF]{standard workload format}
    \newacro{api}[API]{application programming interface}
\end{acronym}
\begin{document}

\title[Improving the Efficiency of a DRL-Based Power Management System for HPC Clusters Using CL]{Improving the Efficiency of a Deep Reinforcement Learning-Based Power Management System for HPC Clusters Using Curriculum Learning}

\author{Thomas Budiarjo}
\affiliation{%
  \institution{Department of Computer Science and Electronics, Universitas Gadjah Mada}
  \city{Yogyakarta}
  \country{Indonesia}
}
\email{thomasbudiarjo@mail.ugm.ac.id}
\orcid{0009-0008-6765-9607}

\author{Santana Yuda Pradata}
\affiliation{%
  \institution{Department of Computer Science and Electronics, Universitas Gadjah Mada}
  \city{Yogyakarta}
  \country{Indonesia}
}
\email{santanayudapradata@mail.ugm.ac.id}
\orcid{0009-0009-0470-8456}

\author{Kadek Gemilang Santiyuda}
\affiliation{%
  \institution{Pascasarjana, Institut Bisnis dan Teknologi Indonesia}
  \city{Denpasar}
  \country{Indonesia}
}
\additionalaffiliation{%
  \institution{Department of Industrial Management, National Taiwan University of Science and Technology}
  \city{Taipei}
  \country{Taiwan}
}
\email{gemilang.santiyuda@instiki.ac.id}
\orcid{0000-0002-4432-6059}

\author{Muhammad Alfian Amrizal}
\authornote{Corresponding Author.}
\affiliation{%
  \institution{Department of Computer Science and Electronics, Universitas Gadjah Mada}
  \city{Yogyakarta}
  \country{Indonesia}
}
\email{muhammad.alfian.amrizal@ugm.ac.id}
\orcid{0000-0003-1124-5137}

\author{Reza Pulungan}
\affiliation{%
  \institution{Department of Computer Science and Electronics, Universitas Gadjah Mada}
  \city{Yogyakarta}
  \country{Indonesia}
}
\email{pulungan@ugm.ac.id}
\orcid{0000-0002-5019-1357}

\author{Hiroyuki Takizawa}
\affiliation{%
  \institution{Cyberscience Center, Tohoku University}
  \city{Sendai}
  \country{Japan}
}
\email{takizawa@tohoku.ac.jp}
\orcid{0000-0003-2858-3140}

\renewcommand{\shortauthors}{T. Budiarjo, S. Y. Pradata, K. G. Santiyuda, M. A. Amrizal, R. Pulungan, and H. Takizawa}

\begin{abstract}

High energy consumption remains a key challenge in high-performance computing (HPC) systems, which often feature hundreds or thousands of nodes drawing substantial power even in idle or standby modes. Although powering down unused nodes can improve energy efficiency, choosing the wrong time to do so can degrade quality of service by delaying job execution. Machine learning, in particular reinforcement learning (RL), has shown promise in determining optimal times to switch nodes on or off. In this study, we enhance the performance of a deep reinforcement learning (DRL) agent for HPC power management by integrating curriculum learning (CL), a training approach that introduces tasks with gradually increasing difficulty. Using the Batsim-py simulation framework, we compare the proposed CL-based agent to both a baseline DRL method (without CL) and the conventional fixed-time timeout strategy.
Experimental results confirm that an easy-to-hard curriculum outperforms other training orders in terms of reducing wasted energy usage. The best agent achieves a 3.73\% energy reduction over the baseline DRL method and a 4.66\% improvement compared to the best timeout configuration (shutdown every 15 minutes of idle time). In addition, it reduces average job waiting time by 9.24\% and maintains a higher job-filling rate, indicating more effective resource utilization. Sensitivity tests across various switch-on durations, power levels, and cluster sizes further reveal the agent’s adaptability to changing system parameters without retraining. These findings demonstrate that curriculum learning can significantly improve DRL-based power management in HPC, balancing energy savings, quality of service, and robustness to diverse configurations.
\end{abstract}



\maketitle

\section{Introduction}

High energy usage is a major issue in operating \ac{hpc} systems \citep{Dayarathna,Amrizal}.
To maintain good \ac{qos}, the nodes of an \ac{hpc} system usually remain on standby mode so that the system is ready to immediately execute incoming jobs \citep{Bridi}. Even in standby mode, these nodes consume substantial energy \citep{Barroso}. As a result, the total energy consumption of the system remains high. Moreover, \ac{hpc} systems often consist of hundreds or even thousands of such nodes, further compounding the overall energy consumption.


Nodes can be turned off if they have been in standby mode for a certain period of time \citep{Hikita,Pinheiro,Chen}. This study refers to this method as a \emph{timeout policy}. However, turning off nodes at inappropriate times can delay job execution. For example, if some nodes are turned off and a new job arrives shortly afterward, enough active nodes may not be available to execute the job immediately. In such cases, the turned-off nodes must be powered back on, which introduces additional delays due to the time required for the nodes to become operational again. To make matters worse, waking up or turning off nodes in large-scale \ac{hpc} systems takes significantly longer than in regular computers; in some cases, it can take tens of minutes \cite{Ohmura}. This prolonged transition time is mainly due to the complexity of \ac{hpc} environments, where turning on/off the nodes involves intricate power sequencing, GPU initialization, parallel file system recovery, and network reconfiguration, all of which contribute to substantial delays \citep{agung2017memory, ahmad2017design, amrizal2012improving, moran2024exploring, ottaviano2024controlpulp}. As job arrivals are dynamic and unpredictable, setting the timeout policy to a fixed time might not be the best solution, and, hence, accurately estimating the appropriate time for turning off/on nodes is crucial for efficient power management in \ac{hpc} systems.

\Ac{rl}, a subfield of \ac{ml}, is particularly well suited for prediction in dynamic environments \cite{Sutton}. The combination of \ac{rl} with \ac{dnn}, known as \ac{drl}, has further improved its effectiveness \citep{lavet}, increasing its popularity for predicting incoming jobs and has been integrated with the job scheduler of \ac{hpc} systems in some literature. Kumar et al. \cite{Kumar} designed a scheduler leveraging \ac{drl} to effectively reduce the average job waiting time. Liang et al. \cite{Liang} introduced a \ac{drl}-based model that not only minimizes the average job waiting time but also promotes fairness between large and small jobs, preventing starvation. Similarly, Fan et al. \cite{Fan} developed a two-level hierarchical neural network-based \ac{drl} agent for handling \ac{fcfs} scheduling with backfilling that can adapt to changes in workload without human intervention. Despite their success, none of these studies has explicitly considered the energy efficiency of \ac{hpc} systems. In the domain of energy efficiency, Khasyah et al. \cite{Fitra} applied \ac{drl} to decide when to turn off standby/idle nodes. Their model, trained with the \ac{a2c} algorithm \cite{Sutton}, demonstrated energy savings in \ac{hpc} systems, outperforming most of the fixed-time timeout policies except the extremely short one, i.e., turning off a node every five minutes of idle time, which is highly impractical in real-case scenarios. However, the \ac{a2c} agent in their study significantly increased job waiting time, thus reducing the system's \ac{qos}. Hence, there is a need for improved methods to enhance the model's performance.

In this study, we extend the work of Khasyah et al. and propose an integration of \ac{cl} \citep{Bengio} to the \ac{drl} agent to further improve its performance, balancing energy savings and the system's \ac{qos}. \ac{cl} is one of the training methods to improve the performance of DRL agents/models. It mimics human learning by training models with problems that gradually increase in complexity. For instance, a model is first trained with a low-difficulty, then with a medium-difficulty, and finally with high-difficulty datasets. The sequence of training datasets, known as the curriculum, can accelerate convergence and improve model performance \citep{Weinshall}. This approach has also been applied successfully in several general scheduling problems \citep{mao2019learning, muller2024reinforcement}. However, its impact on \ac{hpc} energy management systems remains unexplored. Consequently, a knowledge gap exists in understanding how CL can improve the performance of a \ac{drl} agent deployed to save energy in HPC systems. Our main goals are twofold:
(1) to improve energy efficiency beyond the existing approach when deploying a \ac{drl} agent to an \ac{hpc} system and (2) to reduce job waiting time.  
In this paper, we incorporate CL with different training strategies across datasets of varying difficulty levels to determine the optimal training sequence. Our contributions are:  
\begin{enumerate}
    \item Developing a systematic approach to generate datasets for training the energy-saving \ac{drl} agent.
    \item Investigating the best curriculum for training the agent based on the generated datasets.
    \item Demonstrating the impact of \ac{cl} on energy reduction and improved \ac{qos} in HPC systems through extensive simulations.
\end{enumerate}


The rest of this paper is organized as follows. Sec.~\ref{sec:Background} discusses a more detailed background and related work of the paper. In Sec.~\ref{sec:ProposedMethod}, we explain the workflow of our proposed methodology, from the methods to generate the datasets for \ac{cl} up to training and testing the \ac{drl} agent.
Sec.~\ref{sec:ExperimentalSettings} explains the experimental settings of this work, which results are interpreted in Sec.~\ref{sec:Results}. Lastly, the conclusions and future work are stated in Sec.~\ref{sec:Conclusions}.




\section{Background}\label{sec:Background}

\subsection{Reinforcement learning}
\ac{rl} is a category within the domain of \ac{ml} that aims to solve problems involving sequential decision-making. When decisions made in earlier stages influence decisions in subsequent stages within an environment, \ac{rl} can be employed to achieve optimal decisions in a continuous manner~\citep{Sutton}.

\begin{figure}[t]
\includegraphics[width=\linewidth]{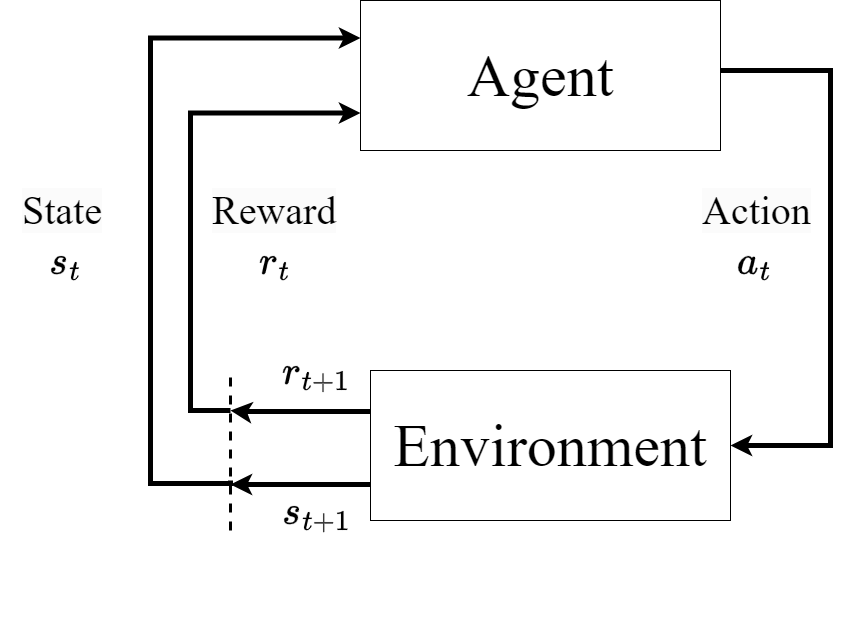}
\centering
\caption{Reinforcement learning~\citep{Sutton}.}
\label{diagramRL}
\end{figure}

In \ac{rl}, an agent learns how to act in an environment to maximize the expected return. The agent's interaction with the environment in \ac{rl} is illustrated in \mbox{Figure~\ref{diagramRL}}. This interaction of an agent and the environment is formulated as a \ac{mdp}. An \ac{mdp} can be described by a tuple $\langle \mathcal{S}, \mathcal{A}, \mathcal{P}, R, \gamma \rangle$, where:
\begin{enumerate}
    \item $\mathcal{S}$ is the set of all possible states.
    \item $\mathcal{A}$ is the set of all possible actions.
    \item $\mathcal{P}$ is the transition probability of the environment's state.
    \item $R$ is the reward function.
    \item $\gamma$ is the reward discount factor.
\end{enumerate}

\begin{figure*}[h]
    \centering
    \includegraphics[width=0.8\linewidth]{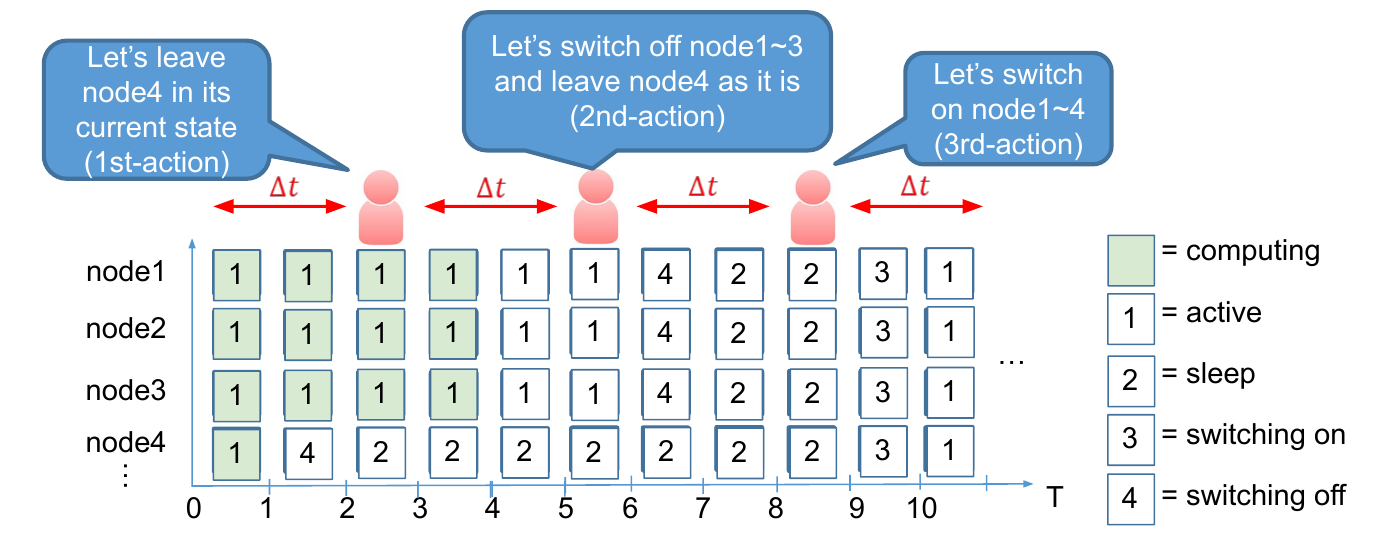}
    \caption{A visualization of the decision-making process in \ac{psmp}.}
    \label{fig:psmp}
\end{figure*}

At timestep $t$, the agent observes the current state of the environment $s_t \in \mathcal{S}$, then decides to take action $a_t \in \mathcal{A}$. The agent takes an action based on the policy $a_t \sim \pi(\cdot\mid s_t, \ldots, s_1)$. Afterward, the environment progresses, returning the state of the next timestep, $s_{t+1}$, and the reward for the agent $r_{t+1}=R(a_t, s_t)$. The primary objective of \ac{mdp} is to find the policy that maximizes the expected return of the environment, $\mathbb{E}_{\pi}[G_t]$, where $G_t$ is given by:
\begin{equation}
    \label{eq:NCF}
        G_t = \sum_{i=0}^{T-t} \gamma^{i}r_{t+i},
\end{equation}
where $T$ is the last timestep and $\gamma \in [0,1]$. The best policy is then denoted as $\pi_* = \underset{\pi}{\arg\max} \; \mathbb{E}_{\pi}[G_0]$.


\ac{drl} is a field that combines deep learning and \ac{rl} \citep{Mnih2015}. Deep learning involves the use of \ac{dnn} to handle complex problems. In \ac{drl}, \ac{dnn}s can be used for representing parts of the agent's policy or embedding the environment state. Various \ac{drl} algorithms have been recently developed and widely used~\cite{Sutton,Mnih,Mnih2015,Schulman17}. One of those is \ac{a2c}, which is a \ac{drl} algorithm that combines two types of models: the actor and the critic. The actor is the part of the agent that makes rules, while the critic provides feedback to improve those rules. \ac{a2c} uses a more accurate estimation of the advantage function to provide feedback to the actor \citep{Mnih}. Specifically, it employs a bootstrapped estimate of the action value rather than relying on the full return, balancing computational efficiency and learning stability. The training steps for \ac{a2c} are as follows:
\begin{enumerate}
    \item Initialize the actor and critic networks with random weights.
    \item Start an epoch with the initial environment and state.
    \item Take and execute an action.
    \item The critic network predicts the advantage value of the performed action, which is the difference between the received reward and the expected reward value of the current state.
    \item Update the weights of the critic network using backpropagation on the critic loss function.
    \item Update the weights of the actor using backpropagation on the actor objective function, which is the average log-likelihood of the actor policy distribution and advantage for each timestep $t$.
    \item Repeat steps 2--6 until convergence or for a specified number of epochs.
\end{enumerate}

\subsection{Power state management problem and its DRL-based solution}
In this section, we briefly present the formulation of a problem called \ac{psmp}, first introduced by Khasyah et al. \cite{Fitra}, along with a \ac{drl}-based method used by the authors to solve it. In \ac{psmp}, a set of computing nodes $M$ of an \ac{hpc} system handles a set of incoming jobs $N$. The incoming jobs first enter the queue before being allocated to the nodes in an \ac{fcfs} manner with backfilling \cite{jiang2009pb}. The job $i \in N$ requests $\mu_i$ nodes to be run, $\mu_i \leq M$, has $e_i$ runtime, and is submitted at time $t^{sub}_i$. A job can be allocated to several nodes, depending on the number of nodes it requests, but a node can only run one job at a time. The starting time of job $i$, $t^{start}_i$, represents the time the job is allocated to the nodes. A running job cannot be canceled, paused, or reallocated. Thus, preemptive job allocation and switching off nodes with a running job are not allowed.


Each node has four possible power states: active, sleep, switching on, and switching off. The power consumption for each state is represented as $p_1$ (active), $p_2$ (sleep), $p_3$ (switching on), and $p_4$ (switching off). Furthermore, the time required for switching on and off is $T_{son}$ and  $T_{soff}$, respectively. A node in the switching-off state must first enter the sleep state before being able to be switched back on again, and vice versa. At time $t$, the power state of node $m\in M$ is determined by the binary flags $(u_{tm1},u_{tm2},u_{tm3},u_{tm4}) \in \{0,1\}^4$. The value of $u_{tm1}$ corresponds to the active state, $u_{tm2}$ the sleep state, $u_{tm3}$ the switching on state, and $u_{tm4}$ the switching off state. The value of $u_{tmn}=1$, $1 \leq n \leq 4$, when node $m$ is in the corresponding state at time $t$, otherwise $u_{tmn}=0$.

The nodes allocation for the job $i$ is denoted by $l_{im}$, where $l_{im}=1$ if job $i$ is allocated to node $m$, and $l_{im}=0$ otherwise. For example, if job $i$ requests two nodes and is allocated to nodes $m=1$ and $m=3$, then $l_{i1}=l_{i3}=1$ and $l_{i\hat{m}}=0, \forall \hat{m}\in M\setminus\{1,3\}$. The computing state of the node $m$ is denoted as $c_{tm}$, where $c_{tm}=1$ if node $m$ is running a job at time $t$, and $c_{tm}=0$ otherwise.

The primary objective of this problem is to determine when and which nodes are switched on and off to minimize energy usage. A simple visualization of such decision-making processes is depicted in Figure~\ref{fig:psmp}. The total energy consumption (in Joules) is formulated as:
\begin{equation}
Z = \sum_{m \in M} \sum_{t=1}^T \sum_{n=1}^4 p_n u_{tmn},
\label{eq:totalPower}
\end{equation}
where $T$ is the time when the whole operation is complete, i.e., when the last job is completed, and each node is either in active or sleep state. Another metric for energy usage is \emph{wasted energy}, defined as the energy consumed when a node is not performing computations. The total wasted energy (in Joules) is formulated as:
\begin{equation}
Z_1 = \sum_{m \in M} \sum_{t=1}^T u_{tm1}(1-c_{tm})p_1 + u_{tm3}p_3 + u_{tm4}p_4.
\label{eq:totalWaste}
\end{equation}
As previously mentioned, we also have to consider the impact of the decision to switch off the nodes on the \ac{qos}, specifically the waiting time of the jobs in the queue. The average waiting time of all jobs, which can be computed when the whole operation is complete, is given by:
\begin{equation}
    Z_2=\sum_{i \in N} \frac{t^{start}_i - t^{sub}_i}{|N|}.
\label{eq:total-waiting-time}
\end{equation}

The wasted energy and waiting time are minimized without violating the constraints:
\begin{align}
    &u_{tm1} \geq l_{im}, \forall i \in N, \forall m \in M, t^{start}_i \leq t \leq t^{start}_i + e_i, \label{eq:constraint:active-computing}\\
    &\delta_{ij} (l_{im} + l_{jm}) \leq 1, \forall m \in M, \forall i, j \in N, i \neq j, \label{eq:constraint:no-double-alloc} \\
    &\sum_{n=1}^4 u_{tmn} = 1, \forall m \in M, 1 \leq t \leq T, \label{eq:constraint:single-state}\\
    &\sum{m \in M} l_{im} = \mu_{i}, \forall i \in N, \label{eq:constraint:num-req-nodes}\\
    &u_{tmn}, l_{im}, c_{tm}, \delta_{ij} \in \{0,1\}, \forall i,j \in N, \forall m \in M, 1 \leq t \leq T, 1 \leq n \leq 4. \label{eq:constraint:range}
\end{align}
Constraint \eqref{eq:constraint:active-computing} ensures that a node must be in active state when computing. Constraint \eqref{eq:constraint:no-double-alloc} ensures that a node cannot run more than one job, where $\delta_{ij}$ is a helper variable, and $\delta_{ij}=1$ if $[t^{start}_i,t^{start}_i+e_i]$ overlaps with $[t^{start}_j,t^{start}_j+e_j]$. Constraint \eqref{eq:constraint:single-state} ensures that every node can only be in one state at a time. Lastly, Constraint \eqref{eq:constraint:num-req-nodes} ensures the number of nodes allocated to job $i$ is exactly as it requests.

We briefly describe the \ac{drl}-based approach to solve \ac{psmp} by first describing the \ac{mdp} formulation of \ac{psmp}. The agent is designed to perform actions to switch nodes on or off at each fixed time interval $\Delta t$, which we refer
as the \textit{action step} and denote by $k$. At the action step $k$, the agent observes the environment state $s_k \in \mathcal{S}$, and performs action $a_k \in \mathcal{A}$. Afterward, the environment progresses and returns the reward $r_{k+1}$ and the next state $s_{k+1}$. Actions are determined by a policy $\pi(\Vec{a}_k \mid s_k)$ that maximizes the expected return $\mathbb{E}_\pi[G_k]$, where $G_k$ is defined as in Eq.~\eqref{eq:NCF}.

The state $s_k$ is defined as the system features $X_k = (\Vec{x}_{k1},\ldots,\Vec{x}_{kM})$, where $\Vec{x}_{km}$ is the feature of the node $m$ before the action $\Vec{a}_k$ is executed. The actions are denoted as $\Vec{a}_k = (a_{k1},\ldots,a_{kM}) \in \mathcal{A}=\{0,1\}^M$. The value $a_{km}=0$ indicates an action of switching node $m$ off at decision step $k$, while $a_{km}=1$ indicates switching it on. The policy used for the decision-making is parametrized by a \ac{dnn} whose parameters are $\Vec{\theta}$, represented as $\pi(\Vec{a}_k \mid s_k; \Vec{\theta})$. Rewards $r_k$ are obtained based on the agent's actions $\Vec{a}_k$.

The reward function combines two components: wasted energy and job waiting time, denoted as $R_1$ and $R_2$, respectively. The $R_1$ reward is expressed as:
\begin{equation}
    R_1(k,k+1) =
    \sum_{m\in M} \sum^{t=\Delta t \times (k+1)}_{t=\Delta t \times k}   u_{tm1}(1-c_{tm})p_1 + u_{tm3}p_3 + u_{tm4}p_4,
\end{equation}
where $R_1(k,k+1)$ computes the wasted energy between the decision steps $k$ and $k+1$. The $R_2$ reward is expressed as:
\begin{equation}
    R_2(k,k+1) = \sum_{i \in N} \sum^{t=\Delta t \times (k+1)}_{t=\Delta t \times k} q_{ti},
\end{equation}
where $R_2(k,k+1)$ the total waiting time of the jobs in the queue between decision steps $k$ and $k+1$, $q_{ti}=1$ if job $i$ is still in the queue at time $t$, and $q_{ti}=0$, otherwise. Combining these components balances energy savings and job waiting time:
\begin{equation}
r_k = -\alpha \frac{R_1 (k,k+1)}{Mp_1\Delta t}- \beta \frac{R_2 (k,k+1)}{J(k,k+1)},
\label{eq:rewardGabung}
\end{equation}
where $0\leq \alpha, \beta \leq 1$, $\alpha+\beta=1$, and they determine the weights of the two reward components. $J(k, k+1)$ represents the possible maximum total job waiting time in the queue during actions $k$ and $k+1$, given by:
\begin{equation}
    J(k,k+1) = \sum_{i\in N} \min\left(1, \sum^{t=\Delta t \times (k+1)}_{t=\Delta t \times k} q_{ti} \right) \Delta t.
\end{equation}


\subsection{Curriculum learning}

\hyphenation{Basic-Shapes}
The concept of \ac{cl} and its application to improve training in \ac{ml} was first formalized by Bengio et al.~\cite{Bengio}. At its core, \ac{cl} is a strategy to train different concepts to a model at different times. The training comprises a sequence of different tasks, commonly starting with easy tasks first, followed by tasks with gradually increasing difficulty. To adopt \ac{cl}, one needs to determine how to design these tasks and their sequence in the training. In addition, one also needs to consider when to switch from one task to the next. The experimental study by Bengio et al.~\cite{Bengio} for shape recognition and language modeling has shown that models trained with \ac{cl} perform better than models trained without \ac{cl}.


Due to its promising results, \ac{cl} has also been adopted to various \ac{drl} studies, e.g., in job-shop scheduling problem (JSSP)~\cite{waubert,muller2024reinforcement}, routing problems~\cite{Ma2021}, autonomous vehicles~\cite{Dinneweth2022}, scheduling in data processing clusters~\cite{mao2019learning}, and job scheduling in \ac{hpc}~\cite{Fan}. Interested readers are referred to a survey of \ac{cl} for \ac{drl}~\cite{Narvekar2020}.

One study of \ac{cl} in \ac{drl} that is closely related to our study is DRAS~\cite{Fan},  a \ac{drl} agent for job scheduling in \ac{hpc}. Interestingly, although the authors did not state explicitly for adopting \ac{cl}, it is apparent that they adopt \ac{cl} in training their agent.
Their training strategy allows the agent to learn gradually, from simple average cases to unseen rare cases. This technique has experimentally performed best compared to other possible training strategies.


In this study, we propose to design a similar training strategy to improve the performance of the trained agent in solving \ac{psmp}. We propose three different types of datasets to train the agent based on a real workload trace of the target \ac{hpc} system. We conduct an experimental study to find the ordering of the training phases that result in the best-performing agent.

\begin{table*}[t]
\caption{NASA dataset.}
\label{tab:dataset_nasa}
\begin{tabular}{lllllllll}
\toprule
Job & Submit Time & Wait Time & Run Time & Proc Alloc & User ID & Group ID & Executable Number & .... \\ 
\midrule
1   & 0           & -1        & 1451     & 128        & 1       & 1        & -1                & ...  \\ 
2   & 1460        & -1        & 3726     & 128        & 1       & 1        & -1                & ...  \\ 
3   & 5198        & -1        & 1067     & 128        & 1       & 1        & -1                & ...  \\ 
... & ...         & ...       & ...      & ...        & ...     &          &                   & ...  \\ 
\bottomrule
\end{tabular}
\end{table*}

\begin{figure}[t]
    \centering
    \includegraphics[width=\linewidth]{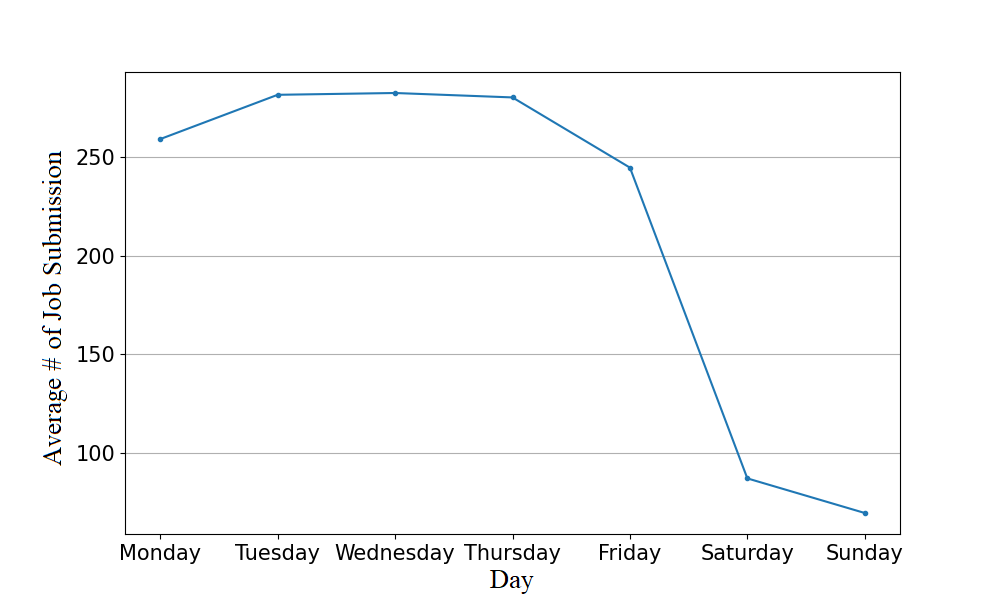}
    \caption{Daily job submission pattern in the real dataset.}
    \label{fig:weeklyGraph}
\end{figure}
\begin{figure}[t]
        \centering
         \includegraphics[width=\linewidth]{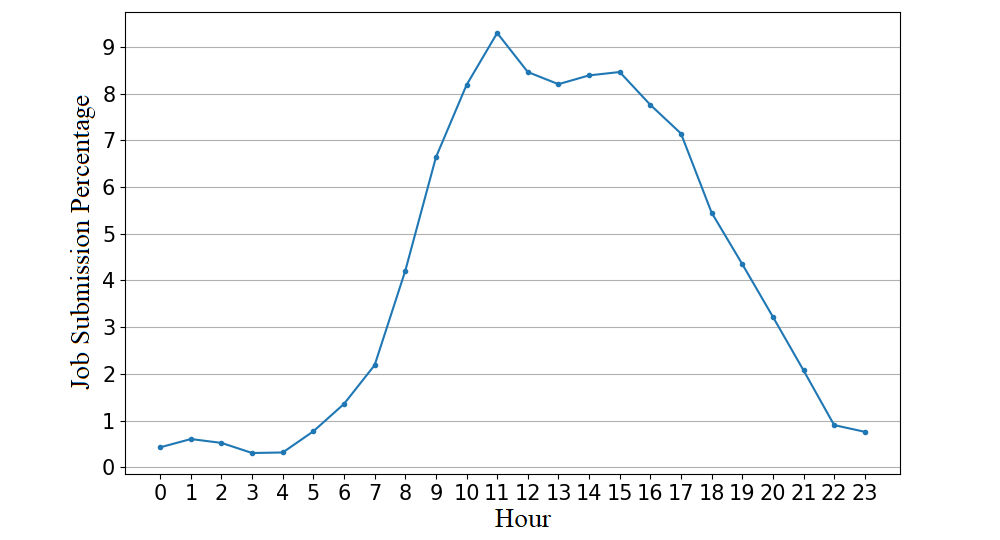}
         \caption{Hourly job submission pattern in the real dataset.}
         \label{fig:dailyGraph}
\end{figure}

\section{Improving the DRL agent with CL to solve PSMP}\label{sec:ProposedMethod}

In this research, we propose to adopt \ac{cl} into the \ac{drl}-based method for \ac{psmp}. To adopt \ac{cl}, a real workload trace of an \ac{hpc} job history is analyzed, and additional datasets are generated. After that, the curriculum sequence is organized based on the generated datasets. The process then continues with training the \ac{drl} model.

\subsection{Dataset description/workload trace}\label{subsec:Dataset}

This study uses a real workload trace from NASA's \ac{hpc} system, accessible via the Parallel Workloads Archive \citep{Feitelson}. The log data contains jobs submitted over three months to the \ac{hpc} iPSC/860 system at NASA Ames Research Center, which has 128 nodes. In this system, each submitted job must request a power-of-two number of nodes---a common requirement in HPC workloads \cite{pollinger2023leveraging}.

This dataset has six available features: Submit Time, Run Time, Number of Allocated Nodes, User ID, Group ID, and Executable Number, as shown in Table \ref{tab:dataset_nasa}. Columns with a value of -1 lack data and cannot be used. Submit Time indicates the time a job is submitted relative to the first submitted job, measured in seconds. Run Time specifies how long the job runs, also measured in seconds. Number of Allocated Processors is the number of nodes allocated to the job. User ID and Group ID represent the IDs of the user and their group, respectively. Additionally, the dataset contains information about StartTime and EndTime, which indicate the date and time when the first job entered the queue.

\begin{figure}[t]
        \centering
         \includegraphics[width=\linewidth]{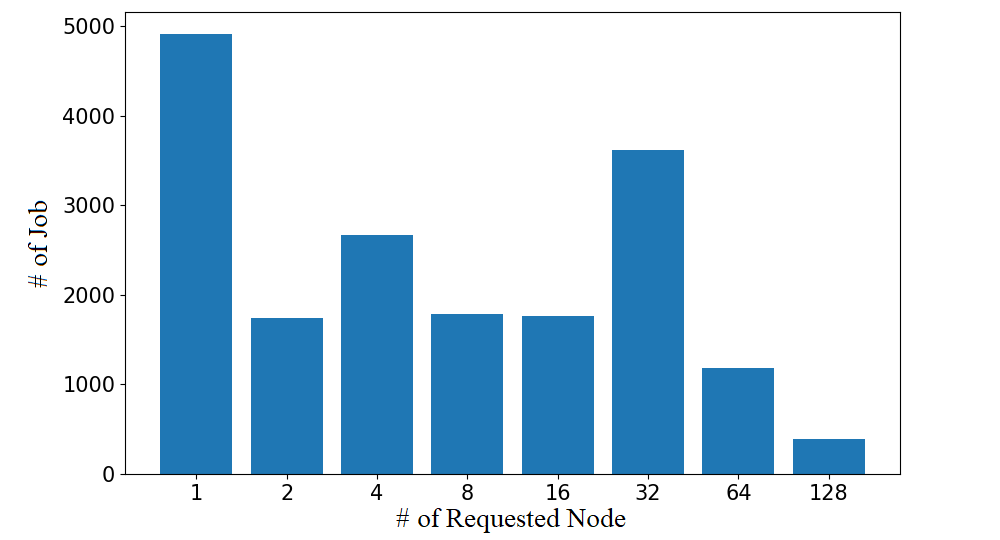}
         \caption{Job size distribution in the real dataset.}
         \label{fig:jobGraph}
\end{figure}
\begin{figure}[t]
        \centering
         \includegraphics[width=\linewidth]{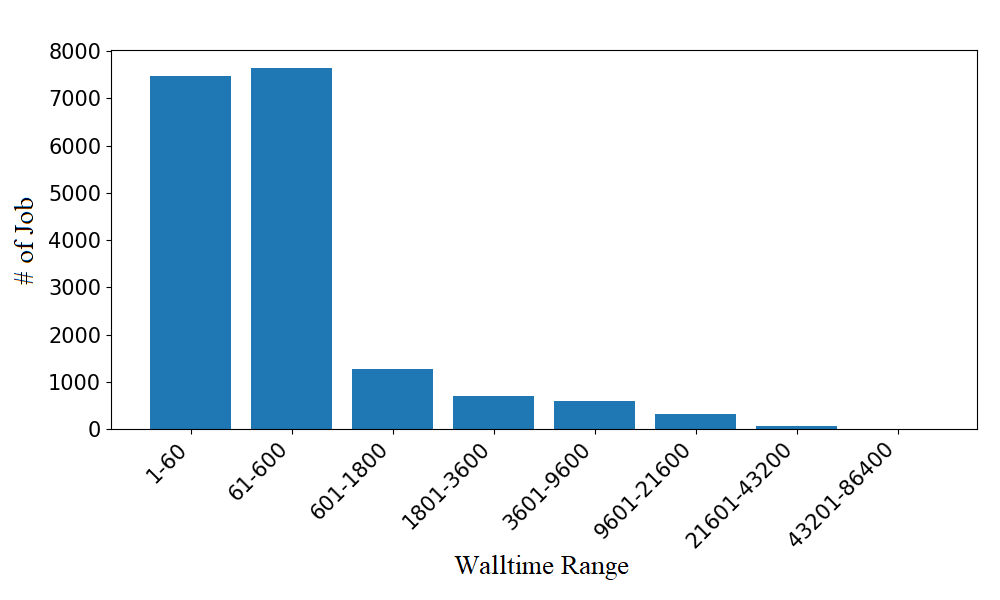}
         \caption{Walltime distribution in the real dataset.}
         \label{fig:walltimeGraph}
\end{figure}

In this research, not all features are required. The features used are Submit Time (Submission Time), Run Time (Walltime), and Number of Allocated Processors (Requested Node). The User ID, Group ID, and Executable Number features are excluded as they are considered less influential on the job runtime. 
The dataset is divided into an 80:20 ratio for training and testing. The dataset used for training is also used to create additional datasets. 
This dataset, referred to as the real dataset, contains 18,067 jobs. Moreover, the statistical patterns of the real dataset---discussed in the following paragraphs---will be used to construct the sampled and synthetic datasets. The arrangements of these three datasets will then be used for the training curricula.

Figure~\ref{fig:weeklyGraph} shows the weekly job submission pattern, where the x-axis represents days, and the y-axis represents the average daily submissions. Figure~\ref{fig:dailyGraph} shows the percentage of jobs submitted by hour, with the x-axis representing hours and the y-axis representing the percentage of total daily submissions. Figure~\ref{fig:jobGraph} displays the job size distribution, where the x-axis represents the number of requested nodes, and the y-axis represents the number of submissions. Meanwhile, Figure~\ref{fig:walltimeGraph} presents the walltime distribution divided into seven groups, with the x-axis representing walltime groups and the y-axis representing the number of submissions.

\subsection{Creating the datasets for CL}

This research employs predefined \ac{cl}, where the difficulty of the dataset is manually determined before training begins. Difficulty indicates how challenging it is for the \ac{drl} agent to predict the arrival patterns of jobs.
In this research, three types of datasets are used: real, sampled, and synthetic datasets.

\subsubsection{Real dataset}

The real dataset consists of raw data that has undergone preprocessing, as explained in Sec.~\ref{subsec:Dataset}. Compared to the sampled dataset, this dataset is categorized as hard difficulty because it contains data originating from real-world scenarios. The real dataset is the source for creating the sampled and synthetic datasets.

\subsubsection{Sampled dataset}
\label{sec:sampleddataset}

The sampled dataset is created by randomly selecting a subset of jobs from the real 
dataset and modifying their submission times to make them easier to predict while 
retaining the characteristics of the real dataset. The submission times are adjusted so 
that the inter-arrival times of this dataset follow an exponential distribution 
commonly used for modeling queues. 
The exponential distribution is memoryless and describes the inter-arrival times in homogeneous Poisson processes; thus, making the job arrival stream more predictable than the real and synthetic datasets. Hence, this dataset can be considered as an easy-difficulty dataset.
To preserve the original dataset's characteristics, the mean inter-arrival time of the real dataset ($\mu$) is calculated by:
\begin{equation}
\label{interarrival}
    \mu = \frac{1}{18066} \sum_{i=1}^{18066} (t_{i+1} - t_i).
\end{equation}
Here, 18066 represents the total number of records in the dataset minus one, and $t_i$ is the submission time of the $i$-th job. Random samples are selected from the entire real dataset as in \cite{Fan}. Then, $\mu$
is used to calculate the rate parameter of the \ac{pdf},
and the submission times of the selected jobs are modified using Algorithm \ref{generatesample}. The function `generateRandomExponential' takes $\mu$ as input and generates random numbers whose \ac{pdf} follow an exponential distribution.

\begin{algorithm}[]
\DontPrintSemicolon
    \caption{Generate Sampled Dataset}
    \label{generatesample}

        \SetKwFunction{FMain}{generateSubmissionTime}
        \SetKwProg{Fn}{Function}{:}{}
        \Fn{\FMain{$F$} \textbf{return} int}{
            interArrivalTime $\gets$ generateRandomExponential(average\_inter\_arrival\_time) \;
            currentTime $\gets$ currentTime + interArrivalTime \;
            \Return currentTime
        }
        currentTime $\gets$ 0 \;
        sampled\_jobs $\gets$ sample from real\_dataset \;
        \For{$i$ \textbf{from} $1$ \textbf{to} length(sampled\_jobs-1)}{
            sampled\_jobs[i]['submission\_time'] $\gets$
            generateSubmissionTime(currentTime)
        }
    
\end{algorithm}

\subsubsection{Synthetic dataset}\label{sec:syntheticdataset}

The synthetic dataset is an artificial dataset designed to resemble the real dataset. As a result, it is classified as a hard-difficulty dataset.
A critical aspect of creating this synthetic dataset is ensuring that it mimics the patterns of the real dataset. This allows the \ac{drl} agent to understand realistic data and improve its adaptability to new data variations. To replicate the real dataset's patterns, the following four types of information are extracted:
\begin{enumerate}
    \item Weekly job submission pattern: the average number of jobs submitted on each day of the week (Monday–Sunday).
    \item Daily job submission pattern: the percentage of jobs submitted within each hour of the day. For instance, jobs submitted between 13:00–14:00 are categorized as jobs for hour 13.
    \item Job size distribution: the distribution of node requests for jobs.
    \item Walltime distribution: the distribution of job walltimes.
\end{enumerate}

Based on the extracted patterns, the StartTime information from the dataset is used to convert submission times into date, day, and time formats. Jobs are generated to match the average daily count from the weekly job submission pattern, and the hourly job counts are derived from the daily job submission pattern. Once each job receives a submission hour, random minutes and seconds are assigned. After all jobs for a day are created, they are sorted by time, and the process continues for the next day. Once all jobs are completed, submission times are converted back to seconds. Walltime and requested nodes are generated based on the walltime and job size distributions in the real dataset. Similar to \cite{Fan}, the synthetic dataset is created in larger quantities than the real dataset to provide more variety for the model to learn from. Therefore, in this study, the synthetic dataset spans one year.

\subsection{Curriculum design}\label{curriculum_design}
The curriculum is created by arranging the sequence of datasets to be trained into the model. Three types of datasets are used: real, sampled, and synthetic datasets, enabling the creation of six different curricula. By combining these three datasets in various sequences, the curriculum can be designed to allow the model to learn progressively, from easy difficulty to hard difficulty or vice versa.

\begin{figure}[t]
\includegraphics[width=\linewidth]{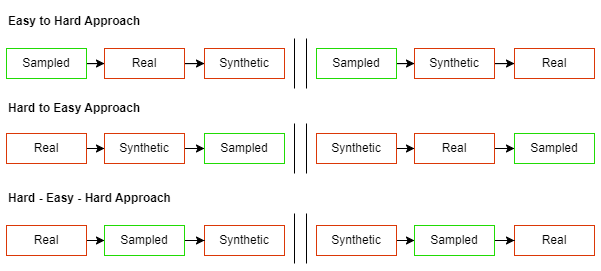}
\centering
\caption{Possible curriculum arrangements.}
\label{fig:curriculum}
\end{figure}

Figure \ref{fig:curriculum} shows three different approaches to curriculum design: easy-to-hard, hard-to-easy, and hard-easy-hard. Each approach has two variations of dataset arrangements. The easy-to-hard approach starts with training using the sampled dataset, followed by the real or synthetic dataset. Conversely, the hard-to-easy approach begins with the real or synthetic dataset and ends with the sampled dataset. In the hard-easy-hard approach, the sampled dataset is flanked by the real and synthetic datasets.
In total, there are six possible curriculum arrangements. In this research,
the impact of all possible arrangements on the model performance is studied.
Therefore, six different \ac{drl} models will be trained, each using one out of six possible curriculum arrangements.

\section{Experimental setup}\label{sec:ExperimentalSettings}


\subsection{Agent training and testing}
\label{sec:AgentTrainingAndTesting}


We set up the agents to operate within a simulated environment of an HPC system. These HPC nodes execute jobs based on the \ac{fcfs} and backfilling scheduling policies. The \ac{fcfs} policy ensures that jobs are executed in the order they arrive, and backfilling is employed to enhance resource utilization. In backfilling, jobs further down the queue may start execution earlier if they do not delay the start of any preceding jobs. This policy effectively fills idle nodes with jobs, minimizing idle time and improving throughput. We implemented these scheduling policies to ensure that the baseline setup of the simulated environment achieves good resource utilization and, hence, the source of the wasted energy is not caused by the scheduler's poor node utilization.

Each node in the environment has distinct power consumption levels depending on its current state. The specific parameter values the agent is trained on is set based on the values used by Khasyah et al.~\cite{Fitra}:
\begin{itemize}
    \item Number of nodes: 128. 
    \item Active and switch-on power consumption: 190 Watts.
    \item Sleep and switch-off power consumption: 9 Watts.
    \item Switch-on time: 45 minutes.
    \item Switch-off time: 30 minutes.
\end{itemize}

The agent uses a set of system features for decision-making, as outlined in Table~\ref{table:features}. Features 1 to 5 represent the overall condition of the \ac{hpc} system and are uniform across all nodes. Conversely, features 6 to 11 are node-specific. These features are updated at every action step to ensure the agent has access to the latest state of the system. The hyperparameter configuration used during training is specified in Table~\ref{tab:hyperparameter}.

All simulations are conducted using the Batsim framework simulator~\cite{dutot2017batsim}, which facilitates the modeling of job scheduling and power management in \ac{hpc} systems. For Python-based interaction, the Batsim-py \ac{api} is utilized~\cite{casagrandebatsim}. The experiments, including model training and job scheduling simulations, are executed on hardware equipped with a 32GB RAM Intel Core i9-9820X CPU @ 3.30GHz (20 CPUs).

\begin{table}[t]
\centering
\caption{The features to represent the state of the system.}
\label{table:features}
\begin{tabular}{cl}
\toprule
\textbf{No.} & \textbf{Features} \\ 
\midrule
1 & Number of jobs in the queue \\ 
2 & Current arrival rate \\ 
3 & Average waiting time of the jobs in the queue \\ 
4 & Total wasted energy \\ 
5 & Average requested runtime of the jobs in the queue \\ 
6 & The power state of node $m$ \\ 
7 & A flag for idling for node $m$ ($u_{tm1}c_{tm}$) \\ 
8 & The current idle time of node $m$\\ 
9 & The release time of node $m$\\ 
10 & The wasted energy of node $m$\\ 
11& The total time node $m$ used for switching on and off\\ 
\bottomrule
\end{tabular}
\end{table}

\begin{table}
\caption{Agent's hyperparameter values.}\label{tab:hyperparameter}
\begin{tabular}{lr}
    \toprule
    \textbf{Hyperparameter} & \textbf{Value} \\
    \midrule
    Number of head in multi-head attention & 8 \\
    Number of layers in multi-head attention & 3 \\
    Embedding size & 128 \\
    Learning rate & $10^{-5}$ \\
    Discounted return $\gamma$ & 0.99 \\
    Gradient clipping norm & 2 \\
    $N_b$ & 64 \\
    $\Delta t$ & 1800 \\
    $\alpha$ & 0.5 \\
    $\beta$ & 0.5 \\
    Training epochs & 10 \\
    \bottomrule
\end{tabular}
\end{table}

\subsection{Comparison methods and evaluation metrics}
\label{sec:evaluation}

In this work, we evaluate the performance of these three approaches on \ac{psmp}: (1) fixed timeout policies, (2) \ac{drl} without \ac{cl} (referred to as the no-CL agent) \cite{Fitra}, and (3) \ac{drl} with our proposed \ac{cl} strategy over six curricula. 
First, the results of training the agent with the six curricula are compared to identify the best outcome, named the \emph{Best Agent}. This Best Agent is compared to the timeout policy methods and the model without \ac{cl}. The timeout policy will be tested using 12 configurations, namely 5, 10, 15, 20, 25, 30, 35, 40, 50, 55, and 60 minutes. The performance of these methods are compared based on
the evaluation metrics given as:
\begin{enumerate}
\item Total wasted energy as formulated in Eq.~\eqref{eq:totalWaste}.
\item Average waiting time as formulated in Eq.~\eqref{eq:total-waiting-time}.
\item Job-filling rate.
\item Number of node shutdowns.
\item The impact on job scheduling performance.
\end{enumerate}
In addition, we evaluate the generalization capability of our Best Agent by testing it in different HPC environments.


\section{Results and discussion}\label{sec:Results}

\subsection{Dataset validation}


In this section, 
we validate and show that our sampled and synthetic datasets follow the desired statistical characteristics. This analysis includes the distribution of job requested nodes, job walltime, daily job submissions, and weekly job submissions, as described in Sec.~\ref{sec:syntheticdataset}. Additionally, the sampled dataset is analyzed to assess its conformity with an exponential distribution.


\subsubsection{Sampled dataset}

\begin{figure}[t]
\centering
\includegraphics[width=\linewidth]{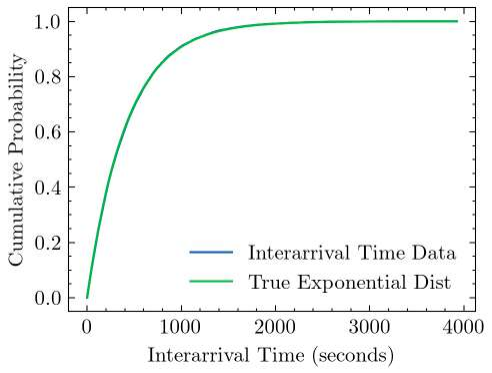}
\caption{
CDF of the inter-arrival time of the sampled dataset compared to the CDF of a true exponential distribution.}
\label{fig:sampleCDF}
\end{figure}





The sampled dataset is a dataset where the walltime and requested node information are sampled from a real dataset. Then, the job arrival stream is generated using an exponential distribution. As a result, the walltime and requested node information will indirectly be similar to the real dataset. 
In Figure~\ref{fig:sampleCDF}, the \ac{cdf} of the inter-arrival time is compared to the \ac{cdf} of the true exponential distribution. The figure shows that the arrival stream of the sampled dataset distribution is extremely close to the true exponential distribution, with a \ac{rmse} of $0.0025$ and a relative \ac{rmse} of $0.0049$ on the sample instances.
This indicates that the sampled dataset follows an exponential distribution; thus, it is more predictable than the real and synthetic datasets---as explained in Sec.~\ref{sec:sampleddataset}.

\subsubsection{Comparison of characteristics of real and synthetic datasets}
\label{hasilReal}

\begin{figure}[t]
\centering
\includegraphics[width=\linewidth]{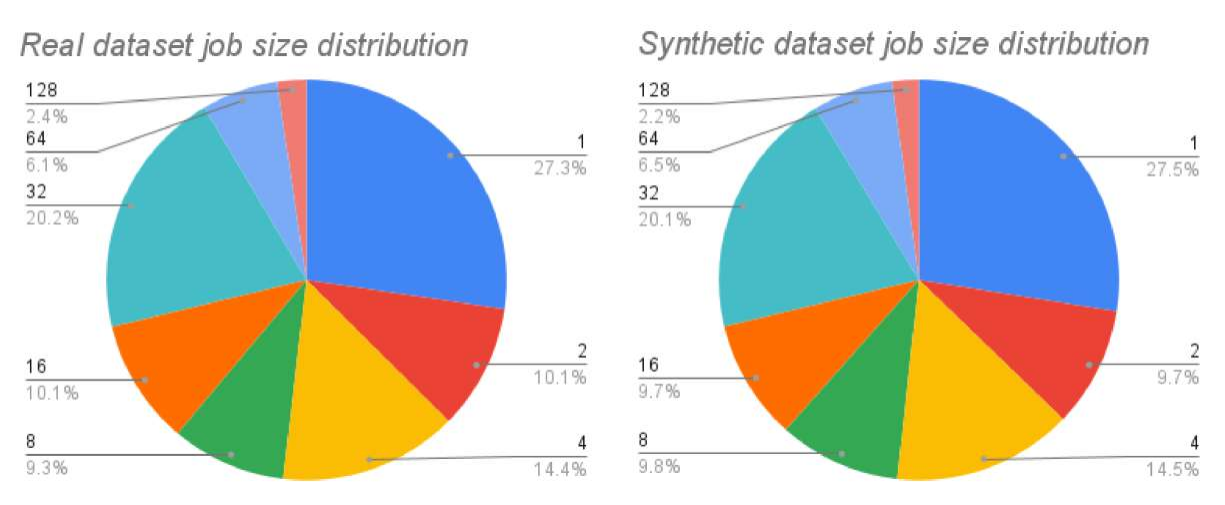}
\caption{Comparison of the requested node distribution.}
\label{fig:jobSizeDistribution}
\end{figure}

\begin{figure}[t]
\centering
\includegraphics[width=\linewidth]{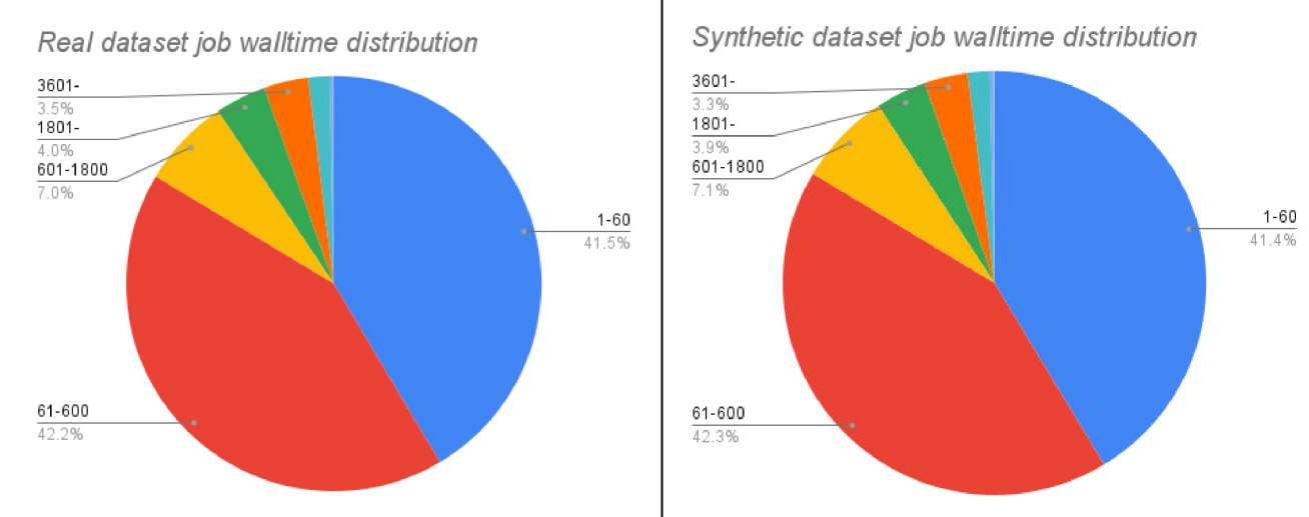}
\caption{Comparison of walltime distribution.}
\label{fig:walltimeDistribution}
\end{figure}

The real dataset serves as a reference for creating additional datasets. 
The distribution of requested nodes in both datasets is shown in Figure~\ref{fig:jobSizeDistribution}. It can be seen that the distribution of requested nodes in the synthetic dataset closely resembles that of the real dataset. Similarly, the distribution of walltime is compared in Figure \ref{fig:walltimeDistribution}, and the distribution of walltime in the synthetic dataset accurately mimics the distribution in the real dataset.

\subsection{Comparison between curricula}


\begin{figure*}
    \centering
    
    \begin{subfigure}{0.96\textwidth}
        \centering
        \fbox{\includegraphics[width=\textwidth]{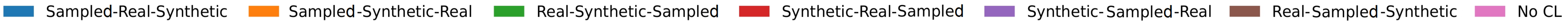}}
    \end{subfigure}
    \begin{subfigure}[b]{0.32\textwidth}
        \centering
        \includegraphics[width=\textwidth]{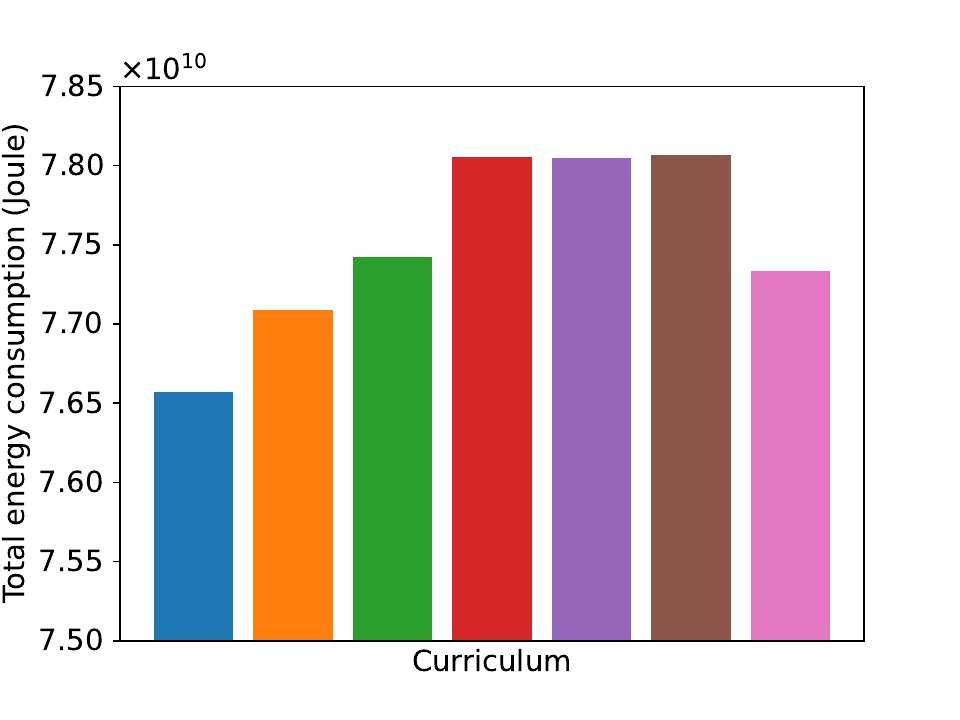}
        \caption{Total energy consumption.}
        \label{fig:total-energy}
    \end{subfigure}
    \begin{subfigure}[b]{0.32\textwidth}
        \centering
        \includegraphics[width=\textwidth]{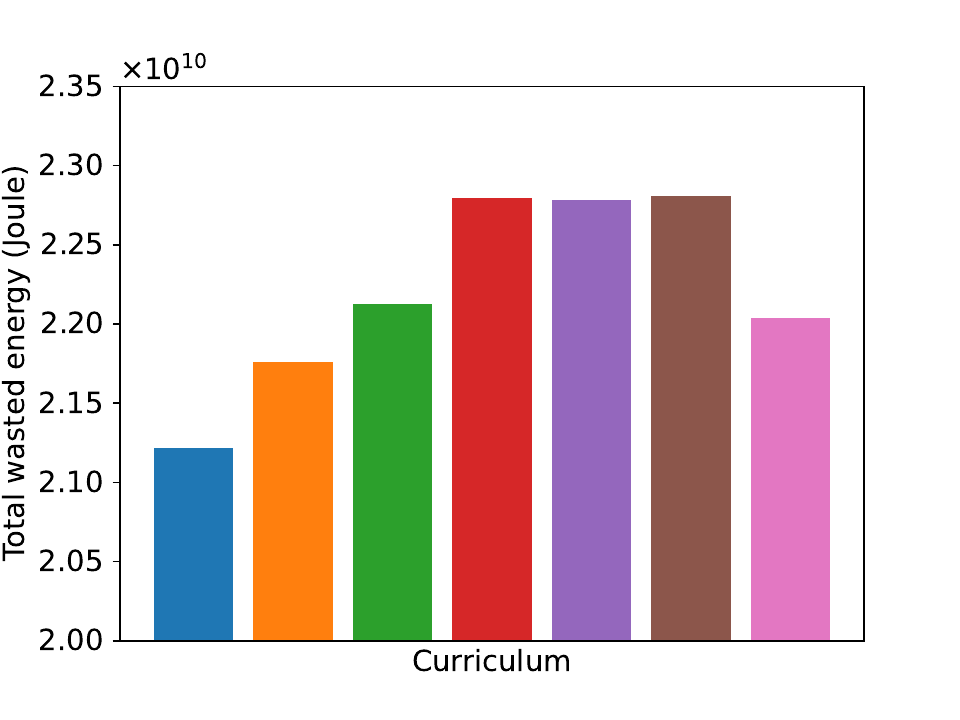}
        \caption{Total wasted energy.}
        \label{fig:waste-energy}
    \end{subfigure}
    \begin{subfigure}[b]{0.32\textwidth}
        \centering
        \includegraphics[width=\textwidth]{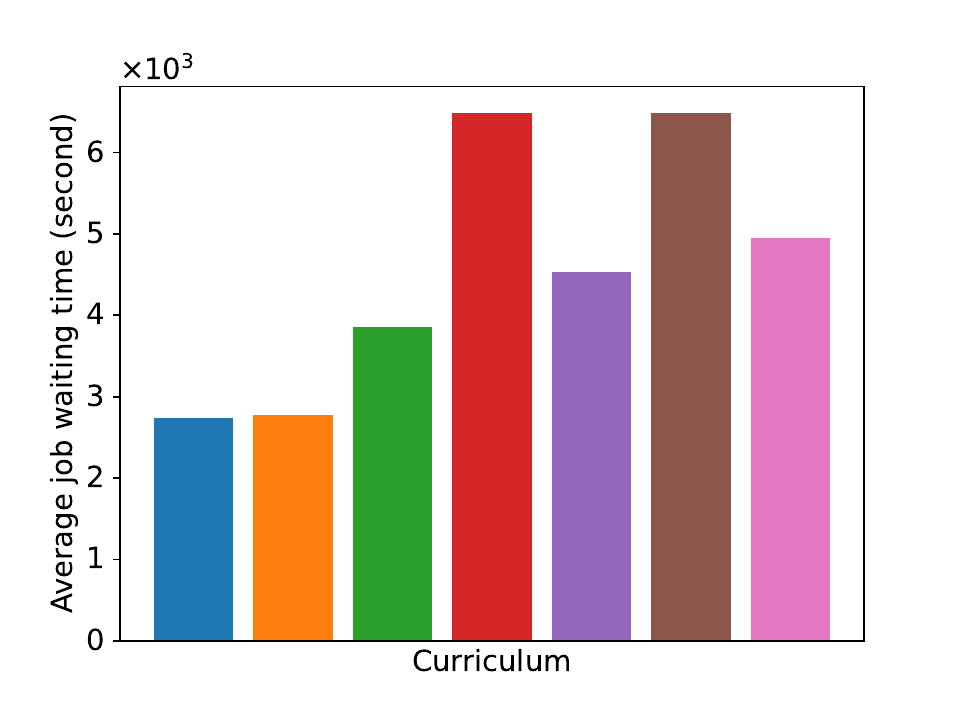}
        \caption{Average job waiting time.}
        \label{fig:avg-waittime}
    \end{subfigure}
    \caption{Evaluation metrics results.}
    \label{fig:multiple_figures}
\end{figure*}


The total energy consumption of the six models trained with different curricula
is shown in Figure~\ref{fig:multiple_figures} (a). The result of the
model trained without \ac{cl} is also included as a reference. As shown by the figure, 
the two models trained by easy-to-hard curricula have lower total energy consumption
than those trained without \ac{cl}, where
the model trained by the sampled-real-synthetic curriculum has the lowest total
energy consumption. On the other hand, the
models trained with hard-to-easy and hard-easy-hard curricula perform
similarly, and even worse, compared to the model trained without \ac{cl}.
To further compare the models' energy-saving performance, we examine
the total wasted energy as depicted in Figure~\ref{fig:multiple_figures} (b).
As shown by the figure, we can see that the difference in total energy consumption
between the models is mainly attributed to the difference in total wasted energy
because the difference in total wasted energy resembles the difference in total energy
consumption. Therefore, we can conclude that the agent trained with sampled-real-synthetic curriculum is the best in terms of energy-saving performance.

In terms of \ac{qos}, i.e., the average job waiting time, the models'
performance is depicted in Figure~\ref{fig:multiple_figures} (c).
The figure demonstrates the consistency of the easy-to-hard curriculum in outperforming other curricula, reducing waiting time by 43.84\% to 44.75\% compared to the no-CL agent configuration. The sampled-real-synthetic sequence consistently outperforms the sampled-synthetic-real sequence by a small margin, making the model trained with the sampled-real-synthetic curriculum the Best Agent in this study.



\subsection{Comparison of the Best Agent with the no-CL agent and other timeout policies}
\label{sec:comparisonagent}

\begin{figure}[t]
    \centering
    \includegraphics[width=\linewidth]{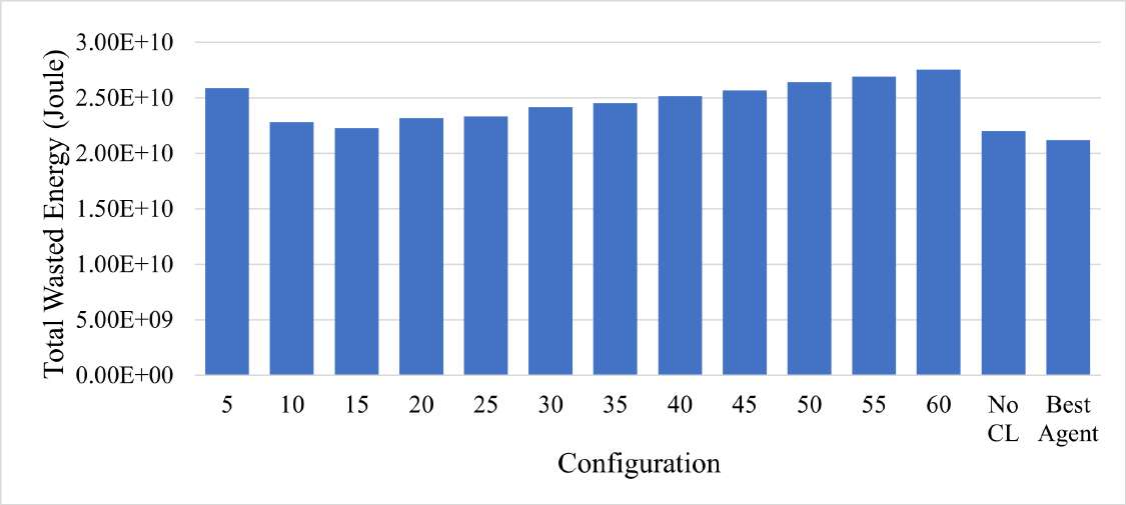}
    \caption{Comparison of total wasted energy for the Best Agent, the no-CL agent, and various timeout policy methods.}
    \label{fig:totalWastedConf}
\end{figure}

This section compares the Best Agent with the no-CL agent and the timeout policy methods. Figure~\ref{fig:totalWastedConf} shows how much total energy is wasted under each approach. From the figure, we can see that the optimal timeout policy configuration is 15 minutes. Our Best Agent achieves a 3.73\% reduction in wasted energy compared to the no-CL agent and a 4.66\% reduction compared to the optimal timeout policy. Although the difference may seem modest in the figure, the Best Agent still saves approximately 800 million Joules more than the no-CL agent—an amount that can translate into substantial cost savings.


\begin{figure}[h]
    \centering
    \includegraphics[width=\linewidth]{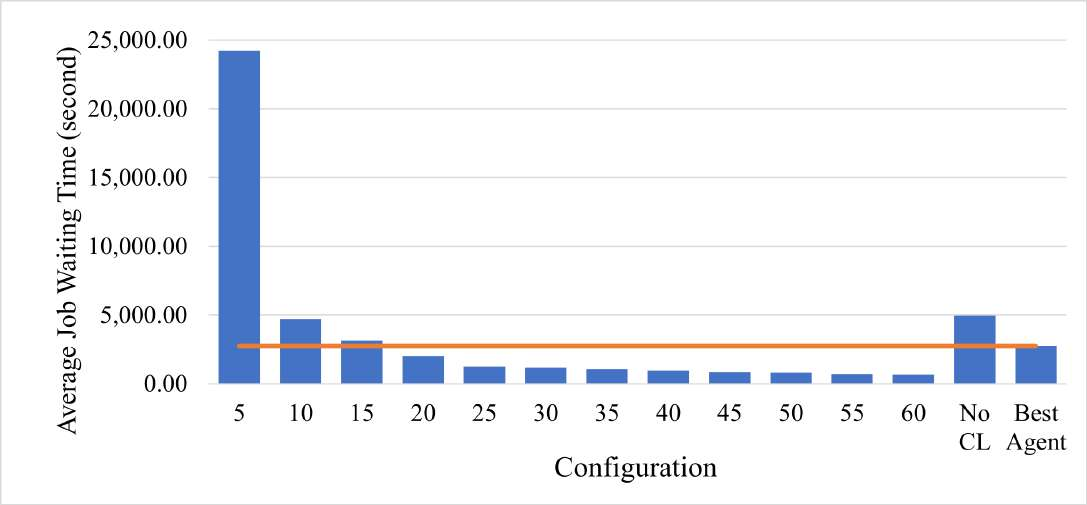}
    \caption{Comparison of average job waiting time for the Best Agent, the no-CL agent, and various timeout policy methods. The orange line indicates the average job waiting time of the Best Agent.}
    \label{fig:averageWaitingConf}
\end{figure}

The Best Agent's advantages extend beyond reducing wasted energy. As shown in Figure~\ref{fig:averageWaitingConf}, the Best Agent achieves a 13.02\% shorter job waiting time compared to the optimal 15-minute timeout policy



\begin{figure}[t]
\centering
\includegraphics[width=\linewidth]{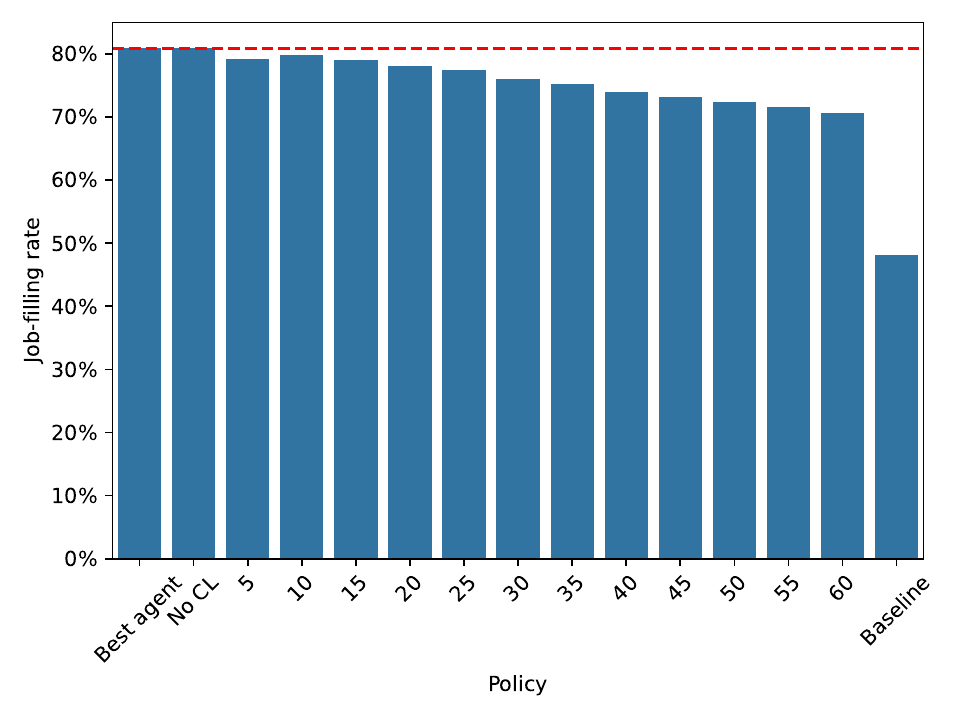}
\caption{The job-filling rate of each strategy on the dataset. The red line indicates the job-filling rate of the Best Agent.}
\label{fig:JobFillingRate}
\end{figure}

To further evaluate the performance of the Best Agent, we analyze the job-filling rate $\eta$ \cite{shoji2017lessons} defined by:
\begin{equation}
\label{jobfillingrate}
    \eta = \frac{t_{compute}}{t_{idle} + t_{compute}},
\end{equation}
where $t_{compute}$ is the total amount of time all nodes spend performing computations, and $t_{idle}$ is the total amount of time all nodes are powered on but not executing any job. This metric reflects how effectively the system is utilized when nodes are active: if a node is powered on but remains idle, the job-filling rate decreases. Figure \ref{fig:JobFillingRate} shows the job-filling rate for each strategy.

In the baseline setting, where nodes remain powered on at all times, the job-filling rate is approximately 48.17\%. This indicates that without turning off idle nodes, the system operates inefficiently. As shown in Figure \ref{fig:JobFillingRate}, our Best Agent and the no-CL agent achieve a notably higher job-filling rate compared to the other policies, surpassing the baseline by 32.70\% and marginally outperforming the optimal 15-minute timeout policy by about 2\%. These results emphasize the effectiveness of RL in minimizing idle time relative to the required computing time, thereby improving overall system utilization. However, as explained in Sec.~\ref{sec:comparisonagent}, our Best Agent still performs better than the no-CL agent and the 15-minute timeout policy in reducing wasted energy and job waiting time.


\begin{figure}[t]
\centering
\includegraphics[width=\linewidth]{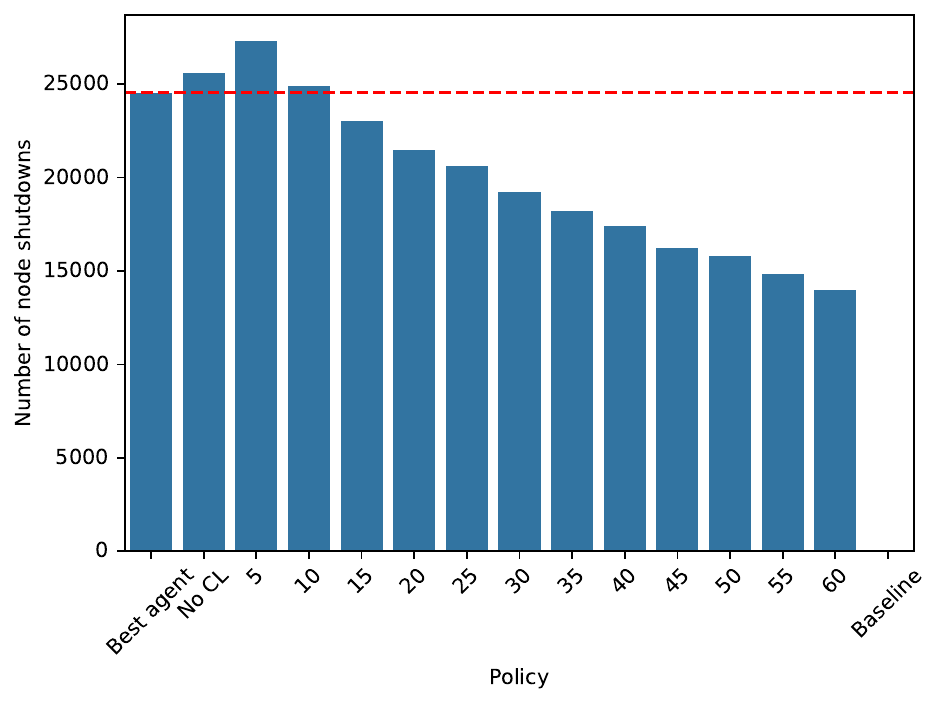}
\caption{The number of node shutdowns using each strategy. The red line indicates the number of nodes shutdown by the Best Agent.}
\label{fig:SwitchOffCount}
\end{figure}


Finally, we evaluate how effectively each strategy manages node power transitions by analyzing the number of node shutdowns. Figure~\ref{fig:SwitchOffCount} illustrates the number of times nodes are powered down under each strategy. In this figure, the number of node shutdowns performed by the Best Agent is between those of the 10- and 15-minute timeout policies, suggesting that, on average, it typically shuts down nodes after an idle interval of 10 to 15 minutes. This aligns with the observation in Figure~\ref{fig:totalWastedConf}, where the 15-minute timeout is shown to be the most energy-efficient among the other timeout policies, while energy consumption starts to increase again with the 10-minute policy. Although the Best Agent still performs shutdowns at such relatively short intervals, its ability to dynamically adjust timing allows it to not only improve energy efficiency but also to help maintain acceptable job waiting times, ensuring a better overall QoS.

\subsection{Comparison of impact on job scheduling}

\begin{figure}[t]
    \centering
    \includegraphics[width=\linewidth]{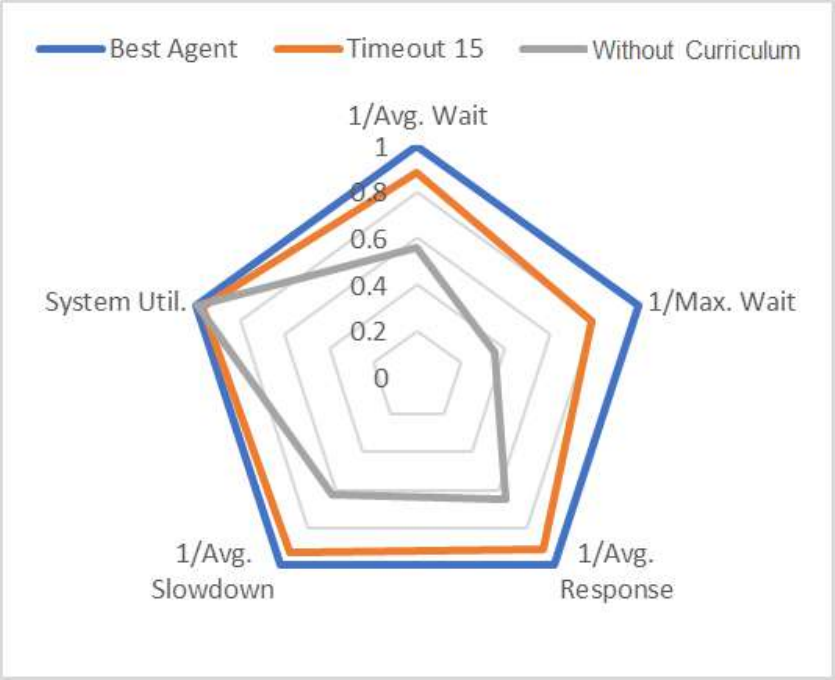}
    \caption{Comparison of impact on scheduling performance.}
    \label{fig:schedulingPerform}
\end{figure}

In this section, we compare
the impact of the Best Agent, no-CL agent, and 15-minute timeout policy on job scheduling-related metrics:  
\begin{enumerate} 
\item Maximum waiting time: The longest time a job waits in the queue. 
\item Average response time: The average time between the job being submitted and completed. 
\item Average slowdown: The average ratio of job response time to the actual time the job is executed. 
\item System utilization: Equivalent to job-filling rate. 
\end{enumerate}
Figure~\ref{fig:schedulingPerform}  shows inverse of the metrics, except for system utilization, and then normalizing them. The larger the area of the graph, the better the overall performance.

The no-CL agent achieves slightly better system utilization than the 15-minute timeout policy, which corresponds to the job-filling rate rankings in Figure \ref{fig:JobFillingRate}. However, it lags behind in all other performance metrics. On the other hand, the Best Agent outperforms both methods, on all metrics.
emphasizing the overall superiority of the Best Agent. 




\begin{figure*}[t]
    \centering
    \includegraphics[width=\linewidth]{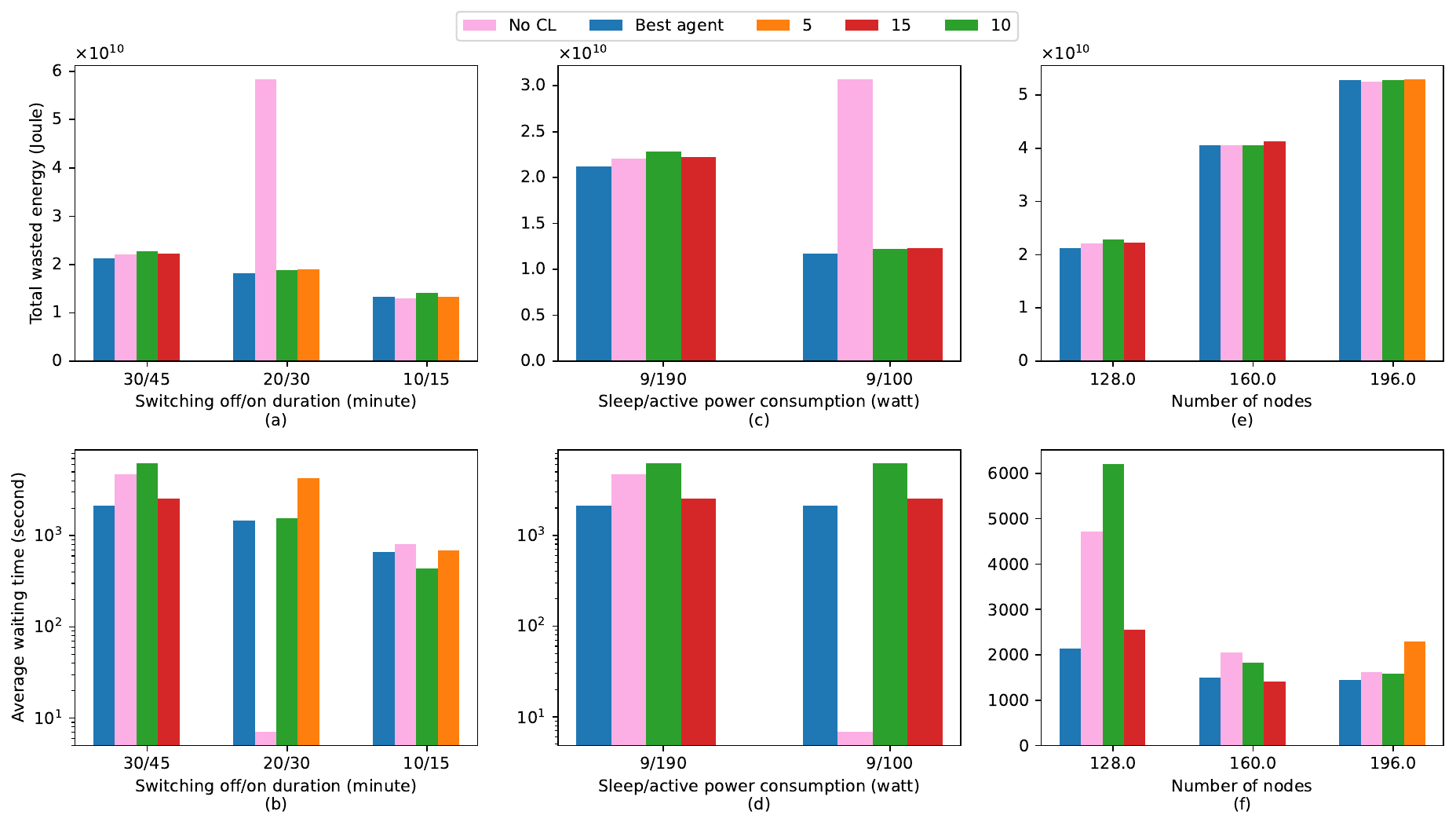}
    \caption{Comparison of average job waiting time and total wasted energy for the Best Agent, the no-CL agent, and two best timeout policies on many variations of switch-on/off duration, sleep/active power consumption, and number of nodes.}
    \label{fig:var}
\end{figure*}

\subsection{Sensitivity study}

We conduct a sensitivity analysis to evaluate how well our model, trained under a specific parameter environment, generalizes to different system configurations. We systematically vary switch-on and switch-off times, power consumption rates, and the number of nodes, respectively. When altering a specific parameter, the others remain fixed at their default values, as described in Sec.~\ref{sec:AgentTrainingAndTesting}. For switch-on/off times, we test three configurations: {30/45, 20/30, 10/15} minutes. To examine the agent's adaptability, we reduce the default 30/45 setting to 20/30 and further down to 10/15, which closely aligns with real-world values observed in Supercomputer AOBA~\cite{ohmura2022toward}. For power consumption, we evaluate two settings: {90/190, 9/100} watts. The second configuration reduces the gap between active and sleep power to test how the agent performs in systems where the power-saving potential is more limited ~\cite{rossi2015impact,georgakoudis2019evaluating}. For the number of nodes, we increase the system size across three configurations: {128, 160, 256}. This expansion in system size enlarges the action space, making decision-making more challenging for the agent and providing insight into its scalability. The results are summarized in Figure \ref{fig:var}. Within each bar group, we compare our Best Agent, the no-CL agent, and the two best-performing timeout policies in terms of wasted energy and average job waiting time for the corresponding parameter settings. We select the two best-performing timeout policies for comparison because their performance fluctuates across different parameter settings.

Figure \ref{fig:var}~(a) and (b) show that the no-CL agent exhibits highly fluctuating performance across different switch-on and switch-off durations. Its occasional promising results appear to be more a matter of luck rather than an indication of true generalization across various scenarios. In contrast, our Best Agent consistently achieves the lowest average job waiting time while maintaining stable performance. In terms of total wasted energy, it performs comparably to the 5-minute timeout policy at switch-on/off times of 10/15 but surpasses all other methods at higher values. This highlights the advantage of CL in enhancing both consistency and generalization across different system configurations.



Figure \ref{fig:var} (c) and (d) highlights once again the inconsistency of the no-CL agent. It performs well in the 9/190 setting, which matches its training environment, but becomes overly aggressive in energy saving at 9/100 power setting. This is evident from its significant reduction in wasted energy, which comes at the cost of an excessive increase in waiting time. These findings indicate that the no-CL agent lacks adaptability and requires retraining to perform well under different conditions—a limitation not observed in our Best Agent. In contrast, our agent significantly outperforms the 10-minute timeout policy in average waiting time at the 9/100 power setting and still maintains an advantage over the 15-minute timeout policy at 9/190. Moreover, it achieves slightly lower total wasted energy than the timeout-based policies. These results demonstrate that our agent remains effective across different switch-on/off times and power consumption rates without requiring retraining.


Figure \ref{fig:var} (e) and (f) reveal that our agent consistently outperforms all other methods in terms of job waiting time while remaining competitive in energy efficiency. Although it is not the most energy-efficient approach, it still achieves reasonable energy savings without compromising the job waiting time. This is an arguably remarkable result, considering that the proposed method was trained for only ten epochs. With additional training or fine-tuning, it is likely that the agent could achieve even better performance, further optimizing both energy efficiency and job scheduling in different node configurations

\section{Conclusion}\label{sec:Conclusions}

Several key conclusions can be drawn based on the research and simulations conducted. Firstly, the easy-to-hard curriculum significantly outperforms the hard-to-easy and hard-easy-hard curricula, with the sampled-real-synthetic training sequence (easy-to-hard) yielding the best overall performance. This finding suggests that a DRL agent for HPC system power management achieves better performance when it begins by tackling simpler problems before progressing to more complex ones. This structured learning approach enhances the agent's capacity to manage the system more effectively.

The Best Agent surpasses both the model without a curriculum and the optimal 15-minute timeout policy in terms of energy-saving metrics. Specifically, the Best Agent achieves 3.73\% more energy savings than the model without a curriculum and 4.66\% more than the optimal 15-minute timeout policy. These results emphasize the importance of CL in optimizing energy consumption for HPC systems. Furthermore, the Best Agent also outperforms the optimal timeout policy in terms of average job waiting time, a metric where the model without a curriculum underperforms. The Best Agent reduces the average job waiting time in the queue by 13.02\% compared to the optimal timeout policy. This improvement highlights the agent's ability to reduce system congestion, a critical factor for efficient HPC scheduling. The Best Agent also demonstrates superior overall performance in its impact on job scheduling efficiency, with improvements observed across all evaluated metrics (for example, attaining higher system utilization and lower average job slowdown than both the no-CL agent and the timeout-based policy). These results highlight the agent's enhanced capacity to identify the optimal timing for turning HPC nodes on and off, further validating the effectiveness of the proposed CL strategy. Finally, the CL-trained agent's advantages persist across different system configurations. In our sensitivity tests, the Best Agent maintained its energy-saving and wait-reduction benefits under varying node switching durations, power consumption rates, and numbers of nodes without requiring retraining. In contrast, the performance of the no-CL agent fluctuated under these changes. This consistency demonstrates the curriculum approach's strong generalization capability in adapting to diverse HPC scenarios.

Several considerations could further improve the proposed CL strategies. One possible improvement is to enhance training by incorporating CL that adjusts difficulty at each epoch. This adaptive approach could better align the learning process with the agent's evolving capabilities. Another avenue for improvement is the development of additional datasets that more effectively capture correlations between requested nodes, walltime, and user IDs. Such datasets would provide the agent with richer training scenarios. Finally, evaluating CL with datasets containing more diverse numbers of jobs could further generalize the approach, ensuring robustness across varying workloads and job patterns. These considerations present opportunities for further refinement of the CL strategy for HPC system power management.

\begin{acks}
This work was partially supported by the Department of Computer Science and Electronics, Universitas Gadjah Mada, under the Publication Funding Year 2025.
\end{acks}

\balance

\bibliographystyle{ACM-Reference-Format}


\end{document}